\documentclass[12pt,singlecolumn,superscriptaddress,nofootinbib]{revtex4}
\usepackage{graphicx}
\usepackage{amssymb} 
\usepackage{url}     
\usepackage{bm}      
\usepackage{color}
\usepackage{float}
\usepackage{natbib}

\usepackage{graphicx}
\usepackage{placeins}

\newcommand{\bise}{Bi$_2$Se$_3$}

\newcommand{\degree}{^\circ}

\begin{document}

\author{Y. Kolumbus}
\affiliation{Racah Institute of Physics, The Hebrew University of Jerusalem, Jerusalem, 9190401 Israel}
\author{A. Zalic}
\affiliation{Racah Institute of Physics, The Hebrew University of Jerusalem, Jerusalem, 9190401 Israel}
\author{N. Fardian-Melamed}
\affiliation{Physical Chemistry Department, The Hebrew University of Jerusalem, Jerusalem 9190401, Israel}
\author{Z. Barkay}
\affiliation{Wolfson Applied Materials Research Center, Tel-Aviv University, Ramat-Aviv, 6997801, Israel.}
\author{D. Rotem}
\affiliation{Physical Chemistry Department, The Hebrew University of Jerusalem, Jerusalem 9190401, Israel}
\author{D. Porath}
\affiliation{Physical Chemistry Department, The Hebrew University of Jerusalem, Jerusalem 9190401, Israel}
\affiliation{Center for Nanoscience and Nanotechnology, The Hebrew University of Jerusalem, Jerusalem 9190401, Israel}
\author{H. Steinberg}
\affiliation{Racah Institute of Physics, The Hebrew University of Jerusalem, Jerusalem, 9190401 Israel}
\affiliation{Center for Nanoscience and Nanotechnology, The Hebrew University of Jerusalem, Jerusalem 9190401, Israel}

\title{Crystallographic orientation errors in mechanical exfoliation}

\begin{abstract}
We evaluate the effect of mechanical exfoliation of van der Waals materials on crystallographic orientations of the resulting flakes. Flakes originating from a single crystal of graphite, whose orientation is confirmed using STM, are studied using facet orientations and electron back-scatter diffraction (EBSD). While facets exhibit a wide distribution of angles after a single round of exfoliation ($\sigma \sim 5^\circ$), EBSD shows that the true crystallographic orientations are more narrowly distributed ($\sigma \sim 1.5^\circ$), and facets have an approximately $3^\circ$ error from the true orientation. Furthermore, we find that the majority of graphite fractures are along armchair lines, and that the cleavage process results in an increase of the zigzag lines portion. Our results place values on the rotation caused by a single round of the exfoliation process, and suggest that when a 1-2 degree precision is necessary, the orientation of a flake can be gauged by the orientation of the macroscopic single crystal from which it was exfoliated.
\end{abstract}

\maketitle

\date{\today}

\section{Introduction}
The advent of the van der Waals transfer method~\cite{hBNsubstrates2010} has made it possible to fabricate a plethora of hybrid structures consisting of mechanically placed exfoliated layers~\cite{geim2013van}. This method allows for non-epitaxial stacks, with freedom of orientation of the constituent flakes - enabling the investigation of relative crystallographic orientations. It is important to control this inter-layer orientation, also called \textit{twist-angle}, as it is a determinant factor in many types of devices: When graphene is stacked over hBN, small twist-angles result in long-range superlattice effects which break the graphene band into minibands~\cite{hunt2013massive}; The twisted bilayer graphene (tBLG) system exhibits non-trivial effects at small twist angles~ \cite{schmidt_superlattice_2014,QuantumHallEffect2012,tBLG2016} and, remarkably, the emergence of a Mott insulator phase~\cite{Cao_Mott_2018} and unconventional superconductivity~\cite{Cao_SC_2018} at the ``magic-angle" of 1.1$\degree$. 

The interface between graphene and other conductors is also of great interest. When placed at proximity to a topological insulator such as \bise, commensurate orientation conditions are expected to result in strong inter-band hybridization~\cite{zhang2014proximity}. Experiments on such hybrids have shown that aligning the flakes results in the emergence of a highly-doped hole Fermi surface at the interface~\cite{Zalic_2017}.  

One possible and widely used approach for fabricating orientation-sensitive devices is identifying orientations according to apparent crystallographic facets of flakes \cite{mishchenko2014twist}. This approach, however, severely limits the choice of flakes for fabrication, which also have other stringent requirements, such as size, geometry and quality, while most graphene sheets do not have straight facet lines; in addition, there is no guarantee that a clear facet line, once found, is indeed parallel to the crystallographic orientation. In fact, facet lines observed in scales of a few nanometers or more may be composed of any sequence of the 6 orientations, and thus may have any apparent shape and direction; however, it has been shown for suspended graphene \cite{kim2011ripping} and for thin anisotropic sheets in the continuum regime \cite{takei2013forbidden}, that lines parallel or close to symmetry orientations are more likely to occur since such tears have a lower energetic cost. 

When fabricating devices consisting of a single material, this problem can be circumvented. For example, exquisite orientation control can be reached in tBLG devices by tearing a single graphene flake and re-stacking the torn pieces~\cite{Cao_Mott_2018, Cao_SC_2018}. The relative orientation of the two pieces is then known since they have originated from a single crystal. Given that many vdW devices are fabricated by exfoliation from an original bulk single crystal, one may expect flakes exfoliated by a single scotch-tape exfoliation step to retain the original orientation of the bulk single crystal~\cite{Park_2010_SEM_exf_graphene}. However, when considering hybrid devices, consisting of different materials,  such methods cannot be applied, and fabrication usually relies on microscopically visible facets. The relation between these facets and the actual crystallographic orientation is, as of today,  unknown, posing an impediment on the study of angle dependent effects in hybrid 2D systems.

In this work we collect statistical data of graphite and WSe$_2$ flake orientation using facet and fracture lines and electron 
back-scatter diffraction (EBSD). Given that graphene tearing dynamics is to date an open question~\cite{sen2010tearing,takei2013forbidden}, and that no clear theoretical model for facet orientation prediction is available, we take a phenomenological approach and measure the effects of single-generation mechanical cleavage followed by deposition on SiO$_2$ on the orientation distribution of the resulting graphite flakes.  

We measure the orientation of the flakes using optical microscopy and EBSD, and find that when graphite flakes are exfoliated from a single crystal bulk piece, their facets exhibit a rather wide distribution, with $\sigma \approx 5\degree$. Facets are, however, a poor proxy for exact crystallographic orientation, and indeed, by mapping such flakes using EBSD, we find that the true crystallographic orientations have a narrower distribution -- with $\sigma \approx 1.5\degree$. The difference is made up by the mismatch between the true crystallographic orientation and the visible straight facets. We hence conclude that any two flakes exfoliated by a single step from a single-crystal bulk are likely to be relatively oriented within 1.5$\degree$. In addition, we find that in exfoliated graphite, most of the facets are closer to the armchair orientation. This was seen both in HOPG and natural graphite.

\begin{figure}[t!]
\begin{centering}
\includegraphics[scale=0.3]{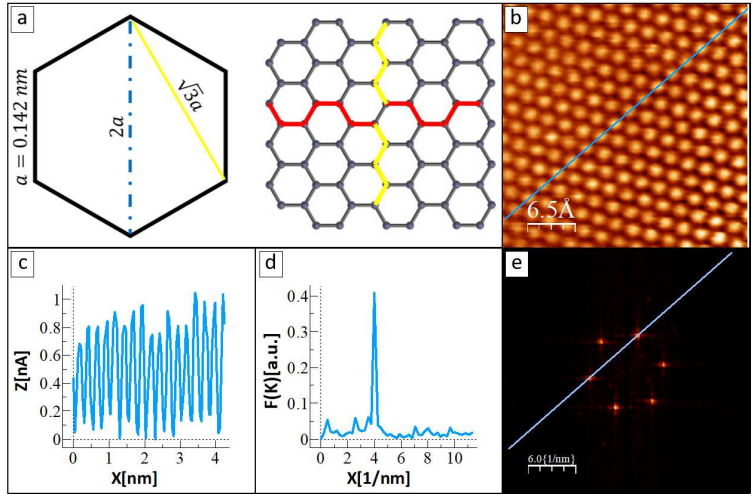}
\end{centering}
\caption{\textbf{(a)} Left: Distances on the hexagon unit; the covalent bond distance is $0.142$~nm, zigzag length unit is $0.256$~nm (yellow line) and the armchair length unit is $0.426$~nm. Right:  Illustration of the two fracture types on the lattice, zigzag in yellow (vertical) and armchair in red (Horizontal). \textbf{(b)} STM scan of the HOPG sample. Note that only half of the atoms are seen as bright spots due to the AB stacking. The blue line is a profile along the zigzag orientation. \textbf{(c)} Periodicity measurement of tunneling current along the zigzag profile. \textbf{(d)} 1D Fourier transform of panel (c). The result of $1/k=0.25\pm0.01$~ nm corresponds to the zigzag spacing. \textbf{(e)} Two dimensional Fourier transform of image b. 
This method gives a direct measurement of both the zigzag and armchair orientations. \label{fig:fig1}}
\end{figure}

\section{Experiments and Results}\label{sec:results}
 
\subsection{STM Measurements}
To find the exact crystallographic orientation prior to exfoliation, we carry out STM measurements of natural graphite and HOPG samples. These crystallographic orientations are later compared to the angular distributions of existing fractures on the surface of the un-exfoliated sample, and to angular distributions of facet lines after exfoliation. 
The materials used in the experiments are from commercially available sources (HQ graphene and MikroMasch for natural graphite and HOPG respectively). The natural graphite is characterized by the vendor using XRD as single crystal, and the HOPG with a mosaic spread is 0.2$\degree$. 
The surface areas of the natural graphite and of the HOPG samples were 25 mm$^2$ and 100 mm$^2$, respectively. The bulk crystals were fixed with silver paste to SiO$_2$ wafers (we used NOVA silicon with 285 nm thermal oxide) and the top layer of each sample was connected directly to the STM drain.  All STM measurements were performed at room temperature at ultra-high vacuum conditions ($\backsim5*10^{-11}$ mbar), using an Omicron LT-STM system. The scans were carried out at constant height mode, with $V_{bias} = 0.7$ V and $I_{set}=1$ nA for the graphite, and $V_{bias}=0.02$ V and $I_{set}=7$ nA for the HOPG. 
Figure 1 shows the orientation results for the HOPG sample, analyzed using two methods: (i) By measuring periodicity along profile lines parallel to the hexagons, we found a periodicity of $0.25\pm0.01$ nm, which corresponds to the zigzag line spacing of $\sqrt{3}*0.142$ nm (Figure 1(b-d)). (ii) By using the 2D Fourier transform (Figure 1(e)). 
Both methods yielded the same crystallographic orientation also when analyzed on multiple locations of each sample. 
The different locations, about 1 mm apart, were scanned by retracting and moving the STM tip, all along preserving the sample orientation as the sample was fixated, where at each spot an area of 5$\times$5 microns was scanned. 
The crystallographic orientations inferred from the STM scans confirm that both the HOPG and the natural graphite samples are mm-scale single crystals or polycrystals at perfect registry. 

\subsection{Fracture Statistics}\label{frac_stat}
We now turn to measuring the distribution of fracture orientations. For the natural graphite we also compare the fracture statistics before and after exfoliation, while in the HOPG case there were almost no fractures on the surface before exfoliation. 
Measurements are performed using optic micrograph mapping of the entire surface of each sample. We use a BX-51 Olympus microscope with $\times$10 magnification at the ocular and $\times$100 objective, yielding $\times$1000 total magnification. 
Then, facet lines are identified and angles are measured as positive angles in the range [$0\degree,180\degree$] with respect to a set reference angle on the SiO$_2$. Total angle measurement error is less than 0.3$\degree$, resulting mainly from image resolution and focus limitations.

\textbf{Natural graphite before exfoliation:} Figure 2(a) shows the fracture statistics of bulk graphite facet orientations. The left panel depicts the raw angular distribution and the right panel is the same data modulo $60\degree$, according to the lattice symmetry. Red lines are Gaussian distribution fits to the data, with a sample size of $141$ fracture lines. The distribution is centered at an armchair direction as obtained from the STM scans (both STM and fracture angles were measured in respect to the same fixed reference angle on the silicon chip).	
Fracture lengths were also measured and no correlation was found between fracture length and its distance from the mean orientation (Pearson correlation test: $\rho=0.0175, p=0.75$).

In view of the different fracture energies between armchair and zigzag, we expect to find a higher occurrence of one type of fracture (though theoretically it is not clear which fracture type should be more frequent). Indeed, in the case of the natural graphite bulk (before exfoliation), $92$\% of the fractures are found close to the armchair orientation, and exhibit a narrow distribution ($\sigma=1.13\degree$). The remaining $8$\% are scattered around the zigzag orientation. This finding is consistent with \cite{incze2001ab} where armchair fracture surfaces were calculated to be more stable than zigzag, and with \cite{kim2011ripping} who found in tearing experiments of suspended graphene that armchair lines were twice as frequent as zigzag lines. This is different than the result of \cite{Girit1705} for the case of hole formation in suspended graphene, where zigzag lines were more prominent (see \cite{GraphenReview2011, GraphenReview2013} for reviews on graphene edges).

\textbf{Natural graphite after exfoliation:} Next, we turn to measure the effect of exfoliation on the distribution of facet angles. Exfoliation is carried out in the following procedure: Adhesive tape is applied to the sample to pull off a thin layer of graphite while keeping the tape connected to a fixed reference; then the tape with the graphite is pressed on to a silicon chip which is aligned with the reference to obtain the reference angle on the new chip, and finally, the tape is removed and the remaining adhered flakes on the chip are examined. We call this method ``first-generation exfoliation" (where the first generation of exfoliated material from the bulk is applied directly to the target substrate). The same procedure is used also for HOPG and for WSe$_2$ (see Section \ref{WSe2}). Most of the flakes created by this first-generation procedure are estimated to be between 10 and 100 nm thick. 

\begin{figure}[t!]
\begin{centering}
\includegraphics[scale=0.35]{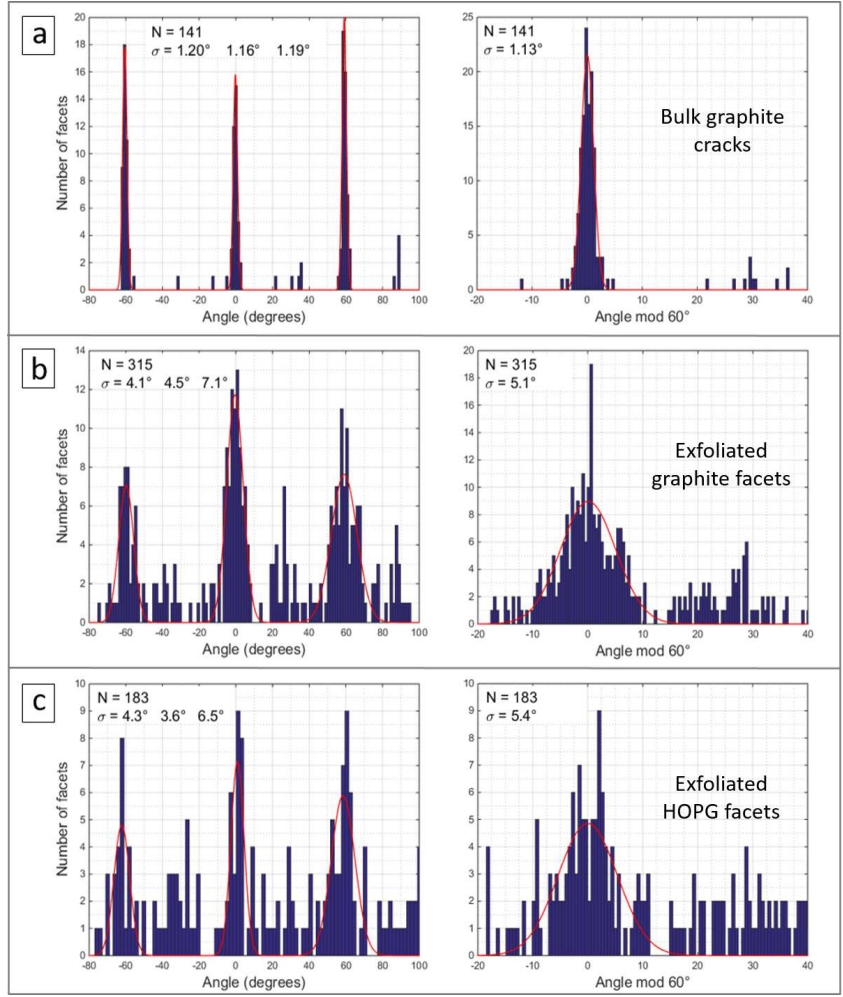}
\end{centering}
\caption{Graphite fracture orientations. Left panels are the raw angular distributions, right panels are the same data modulo $60\degree$. \textbf{(a)} Natural graphite before exfoliation; $92\%$ of the lines are distributed around armchair orientations. Red curves are Gaussian fits with standard deviations of (left to right): $1.20\degree, 1.16\degree, 1.19\degree$. Standard deviation around the armchair angle (right panel) is $1.13\degree$. Sample size: 141 lines. \textbf{(b)} Fractures of the same graphite after exfoliation; $75\%$ of the lines are distributed near armchair directions with standard deviations of $4.1\degree, 4.5\degree, 7.1\degree$. Standard deviation around the armchair orientation is $5.1\degree$. Sample size: $315$ lines. \textbf{(c)} Orientations of exfoliated HOPG; $64\%$ of the lines are distributed near armchair directions with standard deviations of $4.3\degree, 3.6\degree, 6.5\degree$.  Standard deviation around the armchair orientation is $5.4\degree$. Sample size: $183$ lines. 
Results for exfoliated HOPG are similar to those of exfoliated natural graphite. \label{fig:fig2}}
\end{figure}


Fracture statistics of the exfoliated graphite are shown in Figure 2(b). The left panel depicts the raw angular distribution while the right panel shows the data modulo $60\degree$. We find that some information of the initial fracture profile is lost in the exfoliation process -- new fractures are created and possibly some of the fractures move; i.e., exfoliation is not completely angle preserving, also when no angle manipulations are attempted. A number of processes might cause flake rotations and creation of new tearing surfaces. These include metric changes of the flexible tape during exfoliation due to stretching; fluid motion of the tape glue in the initial adhesion stage and local forces which act during separation of the tape from the substrate surface. In addition to the increase in the spread of facet orientations, we also find an increase in the portion of zigzag lines. One generation of exfoliation increased this portion from $8$\% of the facets observed before exfoliation to $25$\% on the exfoliated sample. The reason for this phenomenon is unknown -- and can be the subject of a future investigation, which may probe if it persists with further exfoliation steps. We note that predicting the prevalence of each facet type following exfoliation remains an open question -- connecting nonlinear tearing mechanics with solid state -- yet is beyond the scope of the present study (seminal works in the field include~\cite{incze2001ab,kim2011ripping,PhysRevLett.105.235502,sen2010tearing} and of \cite{takei2013forbidden}).

\begin{figure}[t!]
\begin{centering}
\includegraphics[scale=0.15]{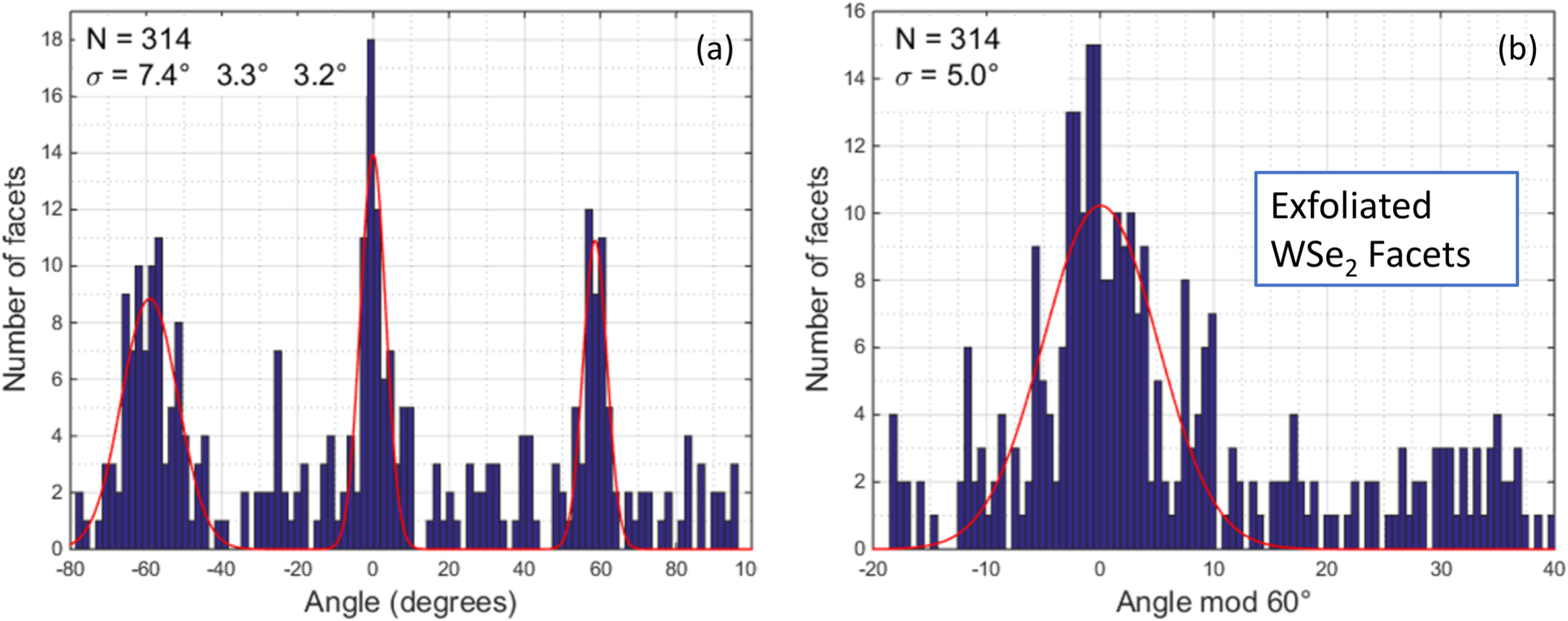}
\end{centering}
\caption{\textbf{a:} Orientations with respect to a reference angle in exfoliated WSe$_2$ after one generation of exfoliation. $70\%$ of the lines are distributed near the crystallographic breaking orientations. Red curves are Gaussian fits with standard deviations of (left to right): $7.4\degree, 3.3\degree, 3.2\degree$. \textbf{b:} the same data modulo $60\degree$. The standard deviation is $5.0\degree$; Sample size: $314$ facets. Results here are similar to the natural graphite and HOPG distributions around the armchair lines, demonstrating that the wide angle distribution is a consequence of the exfoliation process, that is not specific for graphite. \label{fig:fig3}}
\end{figure}
\subsection{Exfoliated WSe$_2$}\label{WSe2}

To test whether the facet orientation distribution is material specific, we have repeated the experiment with the TMD WSe$_2$, which is used ubiquitously in photonic devices~\cite{Lee_2014} and is also useful as a tunnel barrier~\cite{Dvir_2018}. 
We used standard p-type single crystal WSe$_2$ from a commercially available source (HQ graphene) 
and measure fracture orientations in exfoliated WSe$_2$ using the same procedure as in graphite. Here we did not carry out STM scans of the original crystal and the distribution center was taken as zero. 
Figure 3 depicts the angular distribution of facet orientations for first-generation exfoliation in WSe$_2$. 
The observed fracture statistics are strikingly similar to those of natural graphite and HOPG, with a similar standard deviation of $5\degree$ (sample size here was 314 facet lines). This similar facet statistics for different materials implies that the orientation distribution is an outcome of the exfoliation process. The distribution does exhibit a possible secondary fracture orientation with a $30\degree$ shift from the main orientation.

\begin{figure}[t]
\begin{centering}
\includegraphics[scale=0.18]{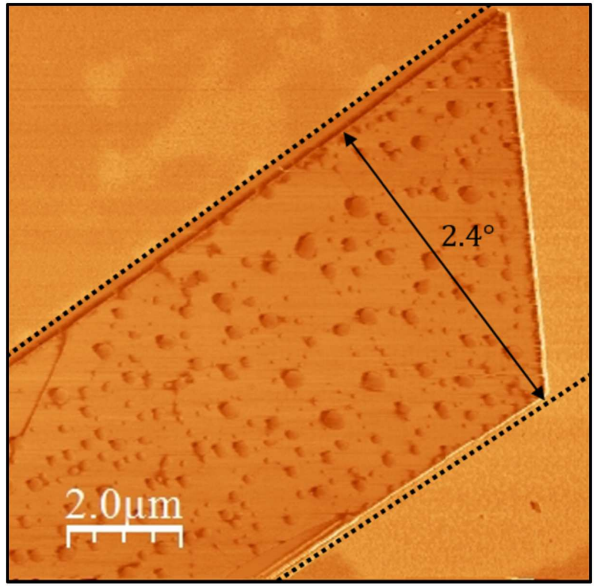}
\end{centering}
\caption{False facets: AFM phase images of a graphite flake. The two seemingly straight facets have a $2.4\degree\pm0.3\degree$ angle between them. \label{fig:fig4}}
\end{figure}
\subsection{False Facets and Local Rotations}\label{fake_facets}
The observed orientation distribution width can either be associated with flake-specific rotations occurring during exfoliation, or with false facets -- lines that are not aligned with the crystal. These may appear to be straight but actually consist of short crystallographic segments. In graphene, specifically, these consist of armchair (zigzag) and small zigzag (armchair) transitions~\cite{Li20061544}. The transition between straight segments may include 5-7 reconstructions~\cite{sen2010tearing}.

STM results indicate that the original flakes before exfoliation are with a single crystallographic orientation, and hence the distribution seen in Figure 2(a) consists of such false facets, while the distributions after exfoliation (Figure 2(b,c)) are a  combination of the two effects. In addition, the theoretical results of \cite{takei2013forbidden} imply that the angle deflection of false facets should be limited, since creation of tearing surfaces with large angle deviations from the orientation is less likely to occur and less likely to create straight lines.
 
Examples of false facets may be seen in atomic force microscopy scans (AFM) of the exfoliated graphite sample. In Figure 4 we show that two facet lines on the same flake, which seem parallel at first sight, actually have a 2.4$\degree\pm$0.3$\degree$ angle between them. We note that using AFM data alone, it is not clear which line is closer to the crystallographic orientation. 

This phenomenon, together with the STM results that show no crystallographic orientation variation at different locations on the sample, imply that false facets are the origin of the distribution width measured on the graphite sample \textit{before} exfoliation (Figure 1(a)). This observation actually places a lower bound on the possible error when trying to estimate graphite orientations based on optically observed lines. That is -- a straight line is not necessarily aligned with the crystallographic orientation. 

\begin{figure}[t]
\begin{centering}
\includegraphics[scale=0.26]{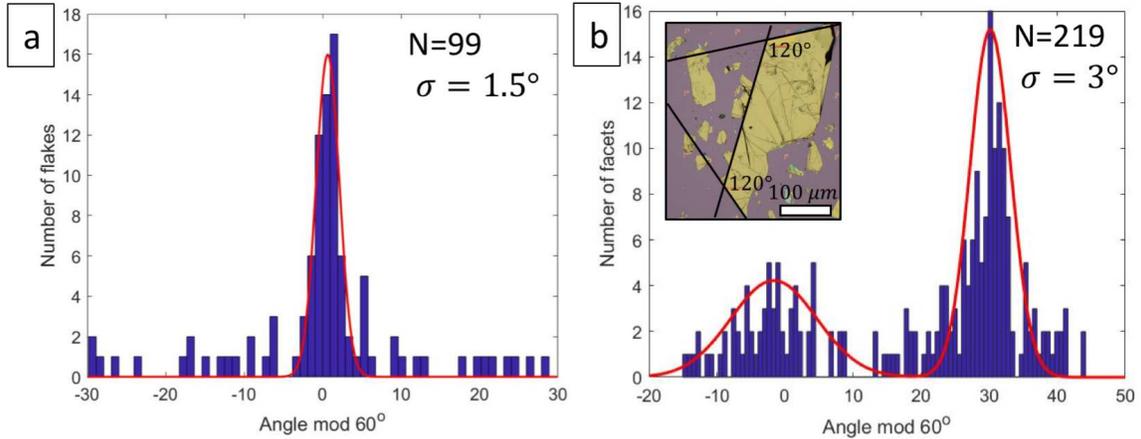}
\end{centering}
\caption{Graphite fracture statistics in comparison to EBSD measure of crystallographic orientation. (a) Crystallographic orientations of 99 flakes as measured by EBSD, with the average angle of the distribution defined as zero. In red, a Gaussian fit giving a standard deviation of 1.5$\degree$. (b) Relative orientations of 219 facets with respect to the EBSD measured orientation of their respective flake. A Gaussian fit to the main peak around 30$\degree$ gives a standard deviation of 3$\degree$. Inset: an example of a graphite flake with three perfectly oriented facets with respect to EBSD measurement. 
 \label{fig:fig5}}
\end{figure}

To differentiate between false-faceting, and actual crystallographic orientation errors incurred upon exfoliation, we utilize EBSD measurements, which allow for a local evaluation of crystallographic orientations with a precision better than 1$\degree$. 
The preparation of the samples follows the same procedure described in Section \ref{frac_stat}. 
We use an Oxford Instruments NORDLYS II EBSD detector, installed in a FEI Quanta 200FEG ESEM, and measurements are taken at two to three points on each flake, with an area map obtained on several sample flakes . The total sample size is 219 flakes  
(see the supplementary material for further details). 
EBSD measurements of exfoliated graphite are presented in Figure \ref{fig:fig5}. 
We thus determine that the true crystallographic orientation distribution width (panel (a)) is $\sigma = $1.5$\degree$. We can then assess the distribution of false facets -- this is done by comparing facet angles with EBSD-derived angles (panel (b)). This error is found to be approximately 3$\degree$. The total observed facet angular distribution is thus associated with both the contribution of actual rotations and with the false-facet angular distribution. 
The distributions we find indicate that when a 1-2--degree precision is required, the orientation of a flake can be estimated by the orientation of the macroscopic single crystal from which it was exfoliated, and that this method has an advantage over the standard method which relies on facets. 

\section{Conclusion}\label{sec:conclusion}
We have demonstrated that upon a single exfoliation step, flakes accumulate a typical 1.5$\degree$ rotation. Facets misrepresent the true crystallographic orientation with typical errors of $\approx$3$\degree$. This poses a challenge for the fabrication of hybrid heterostructures, where existing angle control methods are not applicable (e.g., the methods used in \cite{hunt2013massive,schmidt_superlattice_2014,QuantumHallEffect2012,tBLG2016,Cao_Mott_2018,Cao_SC_2018}). 
Specifically, a hybrid heterosctructure fabricated using facets alone for gauging the orientation will yield a typical relative orientation greater than $3\degree$. Alternatively, it may be possible to improve orientation retention upon the exfoliation process: The $1.5\degree$ rotation we find could be related to either of: 1. minor stretching of the tape, 2. fluid motion of the glue during adhesion, 3. strong local torques during exfoliation. Improving any of these processes could ultimately yield highly oriented flakes, which reflect the orientation of the original bulk.

It is also interesting to test these effects under multiple exfoliation generations and to check correlation to flake thickness. In graphite, we found that the majority of facets are close to the armchair orientation. This distribution may change after additional generations of exfoliation, or after exfoliation in different chemical conditions. It is a challenge for future work to theoretically explain and predict the occurrence frequency of each facet type as a result of exfoliation.

\section*{Acknowledgments}
We thank Prof. Eran Sharon, Prof. Eytan Katzav, Dr. Inna Popov, Prof. Jay Fineberg, Prof. Oded Millo and Tom Dvir, for fruitful discussions and for sharing your advice and knowledge. Funding for our research is provided by ERC-2014-STG Grant No. 637298 and ISF Grant 1363/15. D.P. acknowledge the Israel Science Foundation (ISF grant 1589/14) and the Minerva Centre for bio-hybrid complex systems. D.P. also thanks the Etta and Paul Schankerman Chair of Molecular Biomedicine.

\end{document}